\documentclass[twocolumn]{svjour3}       
\usepackage{graphicx}
\usepackage{cite}
\usepackage{picinpar}
\usepackage{amsmath}
\usepackage{stfloats}
\usepackage{url}
\usepackage{subfigure}
\usepackage{flushend}
\usepackage[utf8]{inputenc}
\usepackage{colortbl}
\usepackage{soul}
\usepackage{multirow}
\usepackage{pifont}
\usepackage{color}
\usepackage{alltt}
\usepackage[hidelinks]{hyperref}
\usepackage{enumerate}
\usepackage{siunitx}
\usepackage{breakurl}
\usepackage{epstopdf}
\usepackage{pbox}
\usepackage{cuted}
\usepackage{mathtools}
\usepackage{color}
\usepackage{wrapfig}
\usepackage{}
\usepackage[]{algorithm2e}
\usepackage[T1]{fontenc}
\usepackage[table]{xcolor}
\usepackage{collcell}
\usepackage{hhline}
\usepackage{pgf}
\usepackage{multirow}

\usepackage{colortbl}
\usepackage{hhline}

\usepackage{amssymb}
\usepackage{changes}

\usepackage{graphicx}
\usepackage{float}
\makeatletter
\newcommand\fs@myRoundBox{\def\@fs@cfont{\bfseries}\let\@fs@capt\floatc@plain
    \def\@fs@pre{\begin{mdframed}[style=myFigureBoxStyle]}%
        \def\@fs@mid{\vspace{\abovecaptionskip}}%
        \def\@fs@post{\end{mdframed}}\let\@fs@iftopcapt\iffalse}
\makeatother

\DeclarePairedDelimiter\floor{\lfloor}{\rfloor}

\begin{document}
    
    \title{Perturbations and Phase Transitions in Swarm Optimization Algorithms}
    
    
    \author{Tom\'{a}\v{s} Vantuch         \and
        Ivan Zelinka \and
        Andrew Adamatzky \and
        Norbert Marwan
    }
    
    
    \institute{Tom\'{a}\v{s} Vantuch \at
        Center ENET at Technical University of Ostrava, Czech Republic \\
        \email{tomas.vantuch@vsb.cz}           
        \and
        Ivan Zelinka \at
        $^1$Modeling Evolutionary Algorithms Simulation and Artificial Intelligence, Faculty of Electrical \&amp; Electronics Engineering, Ton Duc Thang University, Ho Chi Minh City, Vietnam.\\
        $^2$Department of Computer Science at Technical University of Ostrava, Czech Republic \\
        \email{ivan.zelinka@tdt.edu.vn, ivan.zelinka@vsb.cz}           
        \and
        Andrew Adamatzky \at
        Unconventional Computing Lab, UWE, Bristol, UK \\
        \email{andrew.adamatzky@uwe.ac.uk}           
        \and
        Norbert Marwan \at
        Potsdam Institute for Climate Impact Research (PIK), Member of the Leibniz Association, Potsdam, Germany \\
        \email{marwan@pik-potsdam.de}           
    }
    
    \maketitle
    
    \begin{abstract}
        Natural systems often exhibit chaotic behavior in their space-time evolution. Systems transiting between chaos and order manifest a potential to compute, as shown with cellular automata and artificial neural networks. We demonstrate that swarm optimization algorithms also exhibit transitions from chaos, analogous to a motion of gas molecules, when particles explore solution space disorderly, to order, when particles follow a leader, similar to molecules propagating along diffusion gradients in liquid solutions of reagents. We analyze these `phase-like' transitions in swarm optimization algorithms using recurrence quantification analysis and Lempel-Ziv complexity estimation. We demonstrate that converging iterations of the optimization algorithms are statistically different from non-converging ones in a view of applied chaos, complexity and predictability estimating indicators. 
        
        An identification of a key factor responsible for the intensity of their phase transition is the main contribution of this paper. We examined an optimization as a process with three variable factors -- an algorithm, number generator and optimization function. More than 9.000 executions of the optimization algorithm revealed that the nature of an applied algorithm itself is the main source of the phase transitions. Some of the algorithms exhibit larger transition-shifting behavior while others perform rather transition-steady computing. These findings might be important for future extensions of these algorithms.
    \end{abstract}
    

    \section{Introduction}
    
    Natural systems often undergo phase transition when performing a computation (as interpreted by humans), e.g. reaction-diffusion chemical systems produce a solid precipitate representing geometrical structures \cite{costello2017calculating}, slime mould transits from a disorderly network of `random scouting' to prolonged filaments of protoplasmic tube connecting source of nutrients \cite{adamatzky2016advances},  `hot ice' computer crystallizes \cite{adamatzky2009hot}. Computation at the phase transition between chaos and order was firstly studied by Crutchfield and Young \cite{crutchfield1988computation}, who proposed measures of complexity characterizing the transition. The ideas were applied to cellular automata by Langton \cite{langton1990computation}: a computation at the edge of chaos occurs due to gliders.  Phase transitions were also demonstrated for a genetic algorithm which falls into a chaotic regime for some initial conditions \cite{mitchell1993revisiting,wright2001cyclic} and network traffic models \cite{ohira1998phase}.

    Algorithmic models of evolutionary based optimization, AI and Artificial Life possess comparable features of the systems with a higher complexity they simulate \cite{zenil2017algorithmic,detrain2006self}. We focus on the behavioral modes: the presence of random or pseudo-random cycling (analogous to gaseous phase state), ordered or stable states (analogous to solid state), or the chaotic oscillations (transitive states). Each of the modes could imply a different level of computational complexity or an algorithm performance as it was revealed on different algorithms \cite{boedecker2012information,bertschinger2004real,kadmon2015transition}. By detecting such modes we can control and dynamically tune the performance of the computational systems. 
    
    
    \sloppy
    A swarm-like behavior has been extensively examined in studies of Zelinka et al.~\cite{zelinka2017novel} where the changing dynamics of an observed algorithm was modelled by a network structure. The relevance between network features and algorithm behavior supported the control mechanism that was able to increase the algorithm performance \cite{tomaszek2016performance}. An extensive empirical review of existing swarm-based algorithms has been brought by Schut \cite{schut2010model} where approaches like collective intelligence, self-organization,
    complex adaptive systems, 
    multi-agent systems, 
    swarm intelligence
    were empirically examined and confronted with their real models which reflected several criteria for development and verification.
    
    \fussy
    Our previous study \cite{vantuch2018phase} revealed the presence of phase transitions in the computation of various swarm intelligence based algorithms. The different phases were observed on parameters estimating complexity and entropy. In the end, it was also statistically proven, that converging phases significantly differs from others in the view of mean analysis on the used parameters.
    
    This study may serve as an advancement of our previous one. Our goal was to extend the collection of examined swarm-based algorithms, to see whether all of them performs similar transitions. On the other hand, we also extended the testing of swarm-based algorithms into other dimensions, like to rank them according to their sensitivity towards the optimization function or random number generator that drives their computation. This test is considered as necessary in order to reveal whether the phase transition occurs due to the chosen optimized function, used number generator or their appearance is clearly based on the nature of the algorithm. Having this knowledge will shape the design of the optimization algorithms towards more transitional behavior or otherwise to the stability.
    
    \section{Theoretical background}

    \subsection{Swarm based optimization}
    
    The optimization algorithms examined in our study are representatives of bio-inspired single-objective optimization algorithms. They iteratively maintain the population of candidates migrating through the searched space. Their current position represents the solution vector $X$ of the optimized problem.
    
    \subsubsection{Particle Swarm Optimization}
    
    Particle swarm optimization (PSO)
    was proposed by Kennedy et al. \cite{kennedy1995particle}. The main characteristic of the algorithm is the combination of the particle's aim towards the global leader and its previous best position~\cite{kennedy1995particle}. The composition of these two stochastically altered directions modifies its current position to find a better optimum of the given function. Several reviews are available on extensions and variations of the algorithms \cite{banks2007review,del2008particle}.
    
    The process of PSO starts with the initial generation of particles population $P_g$ where $g$ is an index of iteration. Initially, particles are distributed randomly in the searched space with a randomly adjusted vector of velocities $V^g$. Through the generations, all the particles in the current generation are evaluated by the given fitness function. The global leader $b^{g}$ for the entire population is found by its fitness, as well as each particle keeps its personal best position $p^{g}$ from his previous steps. Based on those two positions, the new velocity vector $V^{g+1}$ for each next particles' move is derived.
    
    \begin{equation}
    \begin{split}
    v_i^{g+1} = wv_i^{g} + c_1r_1(b_i^g - x_i^g) + c_2r_2(p_i^g - x_i^g)
    \end{split}
    \label{eq:pso}
    \end{equation}
    where $c_1$ and $c_2$ are the positive acceleration constants, $r_1$ and $r_2$ represent the randomly adjusted variables from the range $\langle 0,1 \rangle$ and $w$ represents the inertia weight from the range $\langle 0,1 \rangle$.
    
    The next generation of particles $P^{g+1}_{PSO}$ is obtained by computing new positions $X^{g+1}$ for each particle accordingly.
    
    \begin{equation}
    x_i^{g+1} = x_i^{g} + v_i^{g+1}
    \end{equation}

    \subsubsection{Self-organizing migrating algorithm}
    (SOMA) is  a stochastic evolutionary algorithm was proposed by Zelinka \cite{zelinka2004soma}, \cite{davendra2016self}. Ideologically, these algorithms stand right between purely swarm optimization driven PSO and evolutionary-like DE. The entire nature of migrating individuals across the search-space is represented by steps in the defined path length and stochastic nature of a perturbation parameter that represents a specific version of the mutation. The randomness is involved through the binary vector by the adjusted perturbation ($PRT$) parameter [0-1] and the given formula
    
    \begin{equation}
    v^{prt}_j = 
    \begin{cases}
    1, & \text{if }  r_j < PRT \\
    0,              & \text{otherwise}
    \end{cases},
    (j=1,2, \cdots, d)
    \end{equation}
    
    Applying $V^{prt}$, the path is perturbed towards a new solution using current particle and leaders position.
    
    \begin{equation}
    \label{cross_soma}
    x^{t+1}_i = x^{t}_i + (x^{t}_L - x^{t}_i)v_i^{prt} 
    \end{equation}
    
    During each migration loop, each particle performs $n$ steps according to the adjusted step size and the path length. If the path length is higher than one, the particle will travel a longer distance, that is his distance towards the leader.

    \subsubsection{Ant colony optimization for continuous domains}
($\text{ACO}_\mathbb{R}$) \cite{socha2008ant} is an extension of an algorithm inspired by ant movements firstly designed to optimize problems in a discrete domain \cite{dorigo2005ant}. This algorithm starts by initialization of the particles' positions at random places in the searched space. These positions, representing the solution candidates, are evaluated according to the optimized function and sorted by their fitness values.

\begin{equation}
f(X_1) \leq f(X_2) \leq f(X_j) \leq f(X_M) 
\end{equation}

From this sorted collection, the weights $w$ are calculated by the form which allows us to prefer solutions with lower fitness values. These may be in a close neighbourhood of the global optimum. Based on the position in the collection, the weights are calculated
\begin{equation}
w_j = \frac{1}{qM\sqrt{2\pi}}e^{-\frac{(j-1)^2}{2q^2M^2}}
\end{equation}
where $q$ is adjustable hyper-parameter controlling the degree on which the lower fitness values are preferred.
The weights are chosen probabilistically towards the leading solution around which a new candidate solution is generated. The
probability of choosing solution $s_j$ as leading solution is given by $w_j/\sum_{a=1}^{k}w_a$ so that the better solutions obtain higher probability to be selected. Once a leading solution $s_{lead}$ is chosen, the algorithm samples the neighbourhood of $i$-th real-valued component of the leading solution $s^i_{lead}$ using a Gaussian PDF with
$\mu_{lead}^i = s^i_{lead}$ and $\sigma_{lead}^i$ is defined as

\begin{equation}
\sigma_{lead}^i = \xi \sum_{j=1}^{k} \frac{|s_j^i - s_{lead}^i|}{k - 1}
\end{equation}

which stands for the average distance between the value of the $i$-th component of $S_{lead}$ and the values of the $i$-th components of the other solutions in the archive, multiplied by a parameter $n$. The process of choosing a guiding solution and generating a candidate solution is repeated in $N$ times (corresponding to the number of 'ants') per iteration. Before the next iteration, the algorithm updates the solution archive keeping only the best $k$ of the $k + N$ solutions that are available after the solution construction process.
    
\subsubsection{Artificial bee colony}

Artificial bee colony (ABC) \cite{karaboga2007powerful} operates with three different kinds of swarm members and with different reaction-diffusion model proposed by Teresenko \cite{tereshko2000reaction,tereshko2002information,tereshko2005collective}. The so called bees are divided into employed, onlooker bees and scout bees. The first group searches for the food around the food source, which computationally means the making use of greedy search over the available solution around the defined position.
\begin{equation}
v_{i,j} = x_{i,j} - \phi_{i,j}(x_{i,j} - x_{k,j})
\end{equation}
where $x_k$ is a randomly selected solution, $j$ is randomly selected index within the dimension of the problem and $\phi_{i,j}$ is a random number within $[-1,1]$. If the value $V_i$ of the fitness is improved, the $x_i$ is substituted by this found position, otherwise $x_i$ is kept.

After all employed bees accomplish their search process, they share their positions with onlooker bees by
\begin{equation}
p_i = \frac{\text{fit}_i}{\sum\limits_{j=1}^{SN} \text{fit}_j}
\end{equation}
where $\text{fit}_n$ is the fitness value of $n$th solution. If the solution is not improved in a defined number of cycles, the food source is being abandoned and scout bees seek for the new source to replace using
\begin{equation}
x_{i,j} = lb_j - \text{rand}(0,1) - (ub_j - lb_j)
\end{equation}
where $\text{rand}(0,1)$ is a generated random number from the normal distribution and $lb$, $ub$ are the lower and upper boundaries of the $j$-th dimension.

\subsubsection{Firefly algorithm}

Firefly algorithm (FA) has been developed in 2008 by Yang and it is based on light flashing interactions of the swarm of so-called fireflies \cite{yang2008luniver,yang2010engineering}. Initially, they are distributed randomly in the searched space which is very similar compared to other swarm-intelligence algorithms. The light flashing interaction represents the algorithm's novelty through the light decay caused by the increasing distance of two interacting flies, and it is defined as follows
\begin{equation}
\beta = \beta_0 e^{-\gamma r^2}
\end{equation}
where $\beta$ is so called the attractiveness, $r$ is the distance and $\beta_0$ is the attractiveness at $r=0$. The attractiveness is estimated based on a current particle's position in the searched space, so it reflects its optimization function value.

The move of the particle is than defined similarly to other optimization algorithms as
\begin{equation}
x_i^{t+1} = x_i^{t} + \beta_0 e^{-\gamma r^2_{ij}}(x_j^{t} - x_i^{t}) + \alpha_t \varepsilon_i^t
\end{equation}
where $x_j^t$ represents the brightest firefly for firefly $x_i^t$ at time $t$ which determines its next move altered by random vector $\varepsilon_i^t$ multiplied by randomization parameter $\alpha_t$ which normally decays over time as
\begin{equation}
\alpha_t = \alpha_0 \delta^t, 0 < \delta < 1
\end{equation}

    \subsection{Number generators driving the process of optimization}
    
    All previously mentioned optimization algorithms more or less rely on a random number generator that adds some controllable amount of stochastic behavior into the process. Altering of its amount may have a critical impact on the convergence which was described and tested in available papers.
    
    Various recent studies showed alternative options able to substitute the random number generator by other mechanisms generating numbers to drive the seek for the global optimum. In studies of Zelinka \cite{zelinka2018investigation}, the chaos number generators proved their quality in the performance increase for various solutions. These studies, therefore, underline the necessity of testing the impact of various number generators on phase transitions of the optimization algorithms.

    \subsection{Complexity estimation}
    
    Three indicators were selected to evaluate the current state of the system represented by swarm-based algorithm. They are the computational complexity derived by Kolmogorov complexity, predictability estimated by the Determinism and the complexity of the deterministic structure in the system represented by an Entropy. Both entropy and determinism are indicators based on recurrence quantification analysis.
    
    \subsubsection{Lempel-Ziv complexity}
    
    According to the Kolmogorov's definition of complexity, the complexity of an examined sequence $X$ is the size of a smallest binary program that produces such sequence \cite{cover2012elements}. Because this definition is way too general and any direct computation is not guaranteed within the finite time \cite{cover2012elements}, approximating techniques are often employed. 
    
    Lempel and Ziv designed a complexity (LZ complexity) estimation in a sense of Kolmogorov's definition, but limiting the estimated program only to two operations: recursive copy and paste \cite{lempel1976complexity}. The entire sequence based on an alphabet $\aleph$ is split into a set of unique words of unequal lengths, which is called a vocabulary. The approximated binary program making use of copy and paste operations on the vocabulary can reconstruct the entire sequence. Based on the size of vocabulary ($c(X)$), the complexity is estimated as 
    \begin{equation}
        C_{LZ}(X) = c(X)(\text{log}_kc(X) + 1)\cdot N^{-1}
    \end{equation}
    where $k$ means the size of the alphabet and $N$ is the length of the input sequence.
    %
    A natural extension for multi-dimensional LZ complexity was proposed in \cite{zozor2005lempel}. In case of a set of $l$ symbolic sequences ${X^i} (i=1,\cdots, l)$, Lempel and Ziv’s definitions remain valid if one extends the alphabet from scalar values $x_k$ to $l$-tuples elements $(x^1_k, \cdots, x^l_k)$. The joined-LZC is then calculated as
    \begin{equation*}
    \begin{aligned}
    C_{LZ}(X^1,\cdots,X^l)= \\ c(X^1,\cdots,X^l)(\text{log}_{k^2}c(X^1,\cdots,X^l) + 1)\cdot N^{-1}.    
    \end{aligned}
    \end{equation*}

    
    Conventionally LZ complexity is used to measure compressibility~\cite{ziv1978compression,feldman1998survey}.  Experimenting with cellular automata we found that the compressibility performs similarly well as Shannon entropy, Simpson index and morphological diversity in detecting phase transitions \cite{redeker2013expressiveness, adamatzky2012diversity}. For example, in cellular automata we can detect formation of travelling localisations, propagating patterns, stable states and cycles~\cite{adamatzky2012phenomenology,ninagawa2014classifying}. The compressibility was also well used for the analysis of living systems, e.g. EEG signals \cite{bhattacharya2000complexity,aboy2006interpretation} and  DNA sequences \cite{orlov2004complexity}, and classification of spike trains~\cite{amigo2004estimating}.

    \subsubsection{Recurrence quantification analysis}

    
    The recurrence plot (RP) is the visualization of the recurrences of $m$-dimen\-sional system states $\vec{x}\in \mathbb{R}^m$ in a phase space \cite{marwan2007recurrence}. Recurrence is defined as closeness of these states $\vec{x}_i\ (i = 1, 2,\ldots,N$ where $N$ is the trajectory length), measured by thresholded pairwise distances. Formally, the RP can be expressed by $R_{i,j} (\varepsilon) = \Theta(\varepsilon- \|\vec{x}_i - \vec{x}_j\|)$ with $\Theta(\cdot)$ the Heaviside step function.
    The Euclidean norm is the most frequently applied distance metric $\|\cdot\|$ and the threshold value $\varepsilon$ can be chosen according to several techniques \cite{koebbe1992use,zbilut2002recurrence,zbilut1992embeddings,marwan2007recurrence,schinkel2008,kraemer2018}.
    
\sloppy
    If only a one-dimensional measurement $u_i$ of the system's dynamics is given, the phase space trajectory has to be reconstructed from the time series $\{u_i\}^N_{i=1}$, e.g., by using the time-delay embedding $\vec{x}_i = (u_i, u_{i+\tau} ,\ldots,u_{i+(m-1)\tau})$, where $m$ is the embedding dimension and $\tau$ is the embedding delay \cite{packard1980geometry}. The parameters $m$ and $\tau$ may be found using methods based on false nearest neighbors and auto-correlation \cite{kantz97}.
    
    \fussy
    
    
    The recurrence quantification (RQA) measures applied in this experiment describe the predictability and level of chaos in the observed system. Determinism is defined as the percentage of points that form diagonal lines
    \begin{equation}
    \label{det_eq}
    DET = \frac{\sum\limits_{l=2}^NlP(l)}{\sum\limits_{l=1}^NlP(l)}
    \end{equation}
    where $P(l)$  is the histogram of the lengths $l$ of the diagonal lines \cite{marwan2007recurrence}. Its values, ranging between zero and one,  estimate the predictability of the system.
    
    %
    %
    The measure divergence is related to the sum of the positive Lyapunov exponents, naturally computing the amount of chaos in the system, and is defined as    
    \begin{equation}
    \label{div_eq}
    DIV = L_{\max}^{-1}, \qquad L_{\max} = \text{max}(\{l_i;i=1,\cdots,N_l\})
    \end{equation}
    where $L_{\max}$ is the longest diagonal line in the RP (excluding the
    main diagonal line)\cite{marwan2007recurrence}.
    

    
    \section{Experiment design}
    
    The motivation is to identify the key factor for the phase transitions in swarm optimization algorithms. \newline Based on our previous study and as it was mentioned previously, we used three metrics $M$ in order to evaluate the phase transitions, they are the Kolmogorov complexity ($m_1 = Kc$), determinism ($m_2 = DET$) and divergence ($m_3 = DIV$). The progress of swarm optimization execution is captured as a tensor $T$ which is part by part ($t_1 \cdots t_N$) examined by metrics $M$. Their changes within one optimization reflected its transitions $\tau$. The intensity of the transition is simply evaluated as the standard deviation of the metric value $\tau_i=\text{std}(m_i(T))$. 
    
    The examined factors that may alter the significance of $\tau$ were represented by the kind of the algorithm ($A$), the number generator ($G \in \{\text{rand}, \text{chaos}, \text{order}\}$) and optimized function ($F$). 
    
    The algorithms were mentioned previously, therefore, $A \in \{\text{SOMA}, \text{PSO}, \text{FA}, \text{ABC}, \text{ACO}_\mathbb{R}\}$. The number of generators were considered as an important source of chaos-order transitions, so three of them were examined ($G \in \{\text{rand}, \text{chaos}, \text{order}\}$). In the first case, the standard random number generator (Mersenne Twister) \cite{matsumoto1998mersenne} was kept to drive the optimization process, while in the $chaos$ and $order$ mechanisms, the numbers were loaded from time series generated by chaotic system -- Lorenz attractor \cite{stewart2000mathematics} (all three coordinates ${x,y,z}$ were used as the source of \textit{randomness}) and deterministic processes -- the $\sin(x)$ equidistantly sampled, similarly as in \cite{zelinka2018investigation}.

    Our aim was to test the algorithms on dimensio\-nally scalable fitness functions $F$ having at least one global optimum surrounded by multiple local optimums. These conditions were met making use of the Rastri\-gin function (Eq.~\ref{rastr_eq}), the Rosenbrock function (Eq.~\ref{rosenb_eq}) and Ackley's function (Eq.~\ref{ackley_eq}) \cite{abiyev2015optimization}, \newline $F \in \{\text{ackley}, \text{rosenbrock}, \text{rastrigin}\}$.
    
    \begin{equation}
    f(x) = A\cdot n + \sum\limits_{i=1}^n (x_i^2 - A\cdot\text{cos}(2\pi x_i))
    \label{rastr_eq}
    \end{equation}

    \begin{equation}
    f(x) = \sum_{i=1}^{N-1} [100(x_{i+1} - x_i^2)^2 + (1 - x_i^2)]
    \label{rosenb_eq}
    \end{equation}
    
    \begin{equation}
    \begin{aligned}
    f(x) = -20\, \text{exp}\,\bigg(-0.2 \sqrt{\frac{1}{N}\sum_{i=1}^{N}x_i^2}\bigg) -\\ \text{exp}\bigg(\frac{1}{N}\sum_{i=1}^{N}\text{cos}(2\pi x_i)\bigg) +20 +e
    \label{ackley_eq}
    \end{aligned}
    \end{equation}

    The entire experiment is, therefore, the set of several executions of the optimization process $O$ which is always defined by those three factors ($O(A, G, F)$). The pseudocode of the experiment may be seen below.
    
\begin{algorithm}
 \KwData{A, G, F, N = 200}
 \KwResult{}
 \For{each a in A}{
  \For{each g in G}{
    \For{each f in F}{
        \For{i in 1 .. N}{     
            $T_i = O(a,g,f)$\;
            \For{each m in M}{
                $\tau_{i,m}=\text{std}(m(T_i))$\;
            }
            save $\tau_i$\;
        }
    }
  }
 }
 \caption{Iterative execution of all adjustments of the experiment}
\end{algorithm}
    
The transitions $\tau$ will differ from each other based on the adjusted factors of optimization. To identify the key factor responsible for the increasing amplitude of $\tau$ we need to compare the means for each factor. An additional outcome will be the reveal of the conditional means for each algorithm while one of its factors will be fixed.

    \subsection{Tensor data obtained from the optimization}
    
    \sloppy
    The optimization step of the optimization algorithms is represented by the positions ($X_{t_1} = \{x_{t_1,1}, x_{t_1,2}, \ldots, \newline x_{t_1,D}\}$) taken by its population members ($P = \{p_1, p_2,$ \newline $\ldots, p_N\}$) during their migrations/iterations ($p_1 = {X_{t_1,1},X_{t_2,1}, \ldots, X_{t_m,1}}$). All of them are stored for the further examination. The time windows $w$ of iterations are taken and transferred into matrices of particles positions where columns are particle's coordinates and rows are ordered particles by their population number and time ($P_{w_{i}} = \{x_{t_i,1}, x_{t_i,2}, \ldots, x_{t_i,N}, x_{t_{i+1},1}, \newline x_{t_{i+1},2}$ $\ldots, x_{t_{i+1},N}, \ldots x_{t_{i+w},N}\}$).
    
    \fussy
%
%

    \begin{figure*}[!tbp]
        \centering
        \subfigure[]{\includegraphics[scale=0.7]{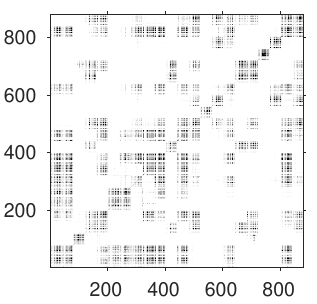}} 
        \subfigure[]{\includegraphics[scale=0.7]{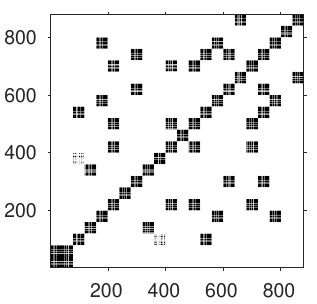}} 
        \subfigure[]{\includegraphics[scale=0.7]{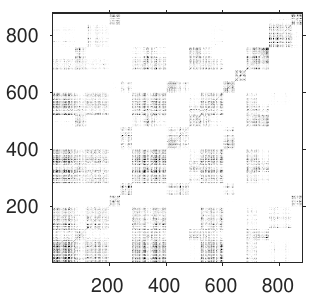}}\\
        \subfigure[]{\includegraphics[scale=0.7]{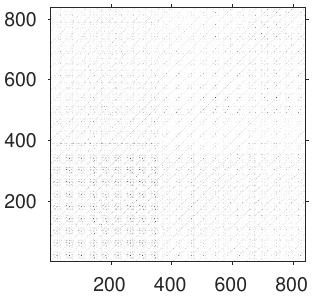}}
        \subfigure[]{\includegraphics[scale=0.7]{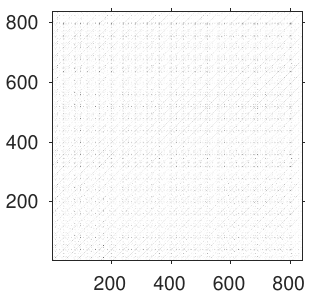}}
        \subfigure[]{\includegraphics[scale=0.7]{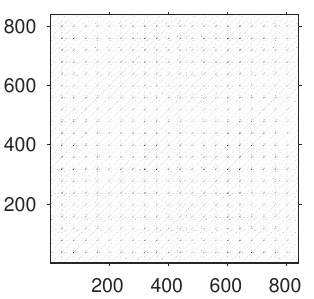}}\\
        \subfigure[]{\includegraphics[scale=0.7]{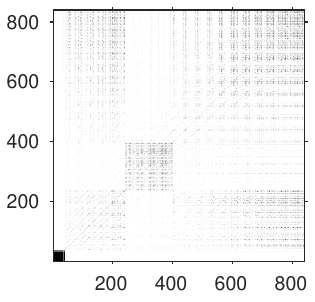}}
        \subfigure[]{\includegraphics[scale=0.7]{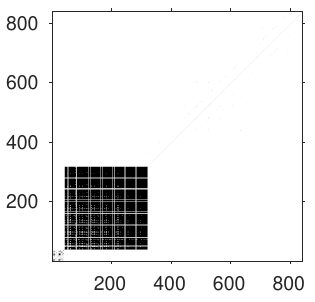}}
        \subfigure[]{\includegraphics[scale=0.7]{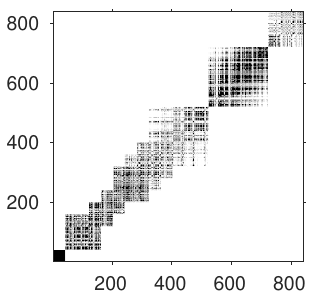}}
        \caption{Recurrence plots of the PSO (abc), DE (def), and SOMA (ghi) behavior calculated as similarities among the particles' positions $X_t$ grouped into the windows of populations $P_{w_i}$ during their (a,d,g)~``post-initial'' (10th migration), 
            (b,e,h)~``top-converging'' (60th migration) and (c,f,i)~``post-converging'' (400th migration) phase.}
        \label{pso_de_soma}
    \end{figure*} 
    
    The changes and interactions inside of their migrating populations are not usually visible in convergence plots; however, changes during the convergence may be estimated using recurrence plots. For this purpose, three selected windows of algorithms' iterations were visualized to spot the differences among them. Figure~\ref{pso_de_soma} illustrates how phases of the algorithm convergences are reflected in RPs.

    \paragraph{Complexity estimation.}
    
    The obtained matrix $P_{w_{i}}$ served as input for a joint Lempel-Ziv complexity (LZC) estimation and RQA.
    
    For the purpose of joint LZC estimation, the input matrix was discretized into adjustable number of letters $n_l$ of an alphabet by the given formula. Let $p_{\min}=\min\{p_j|1 \leq j \leq w\}$, $p_{\max}=\max\{p_j|1 \leq j \leq w\}$ and $p_d=p_{max}-p_{min}$ then each element $p_j$ is assigned value $p_j \leftarrow \floor{n_l \frac{p_j-p_{min}}{p_d}}$.
    %
    The joint-LZC therefore stands, in our case, for the complexity of time ordered $n$ dimensional tuples (populations).
    
    In case of RQA, there is a possibility to directly use the spatial data representation \cite{marwan2007generalised}, therefore we did not apply the Takens' embedding theorem \cite{takens1981detecting,marwan2015} and we directly calculated the RP from our source data. The RQA features like determinism and divergence were calculated.

    \begin{figure*}[!tbp]
        \centering
        \subfigure[]{
            \includegraphics[scale=0.325]{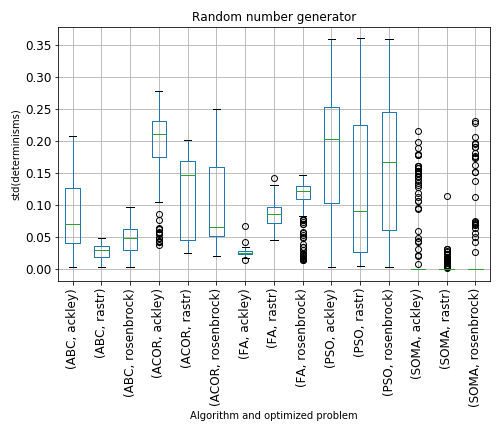}
            \includegraphics[scale=0.325]{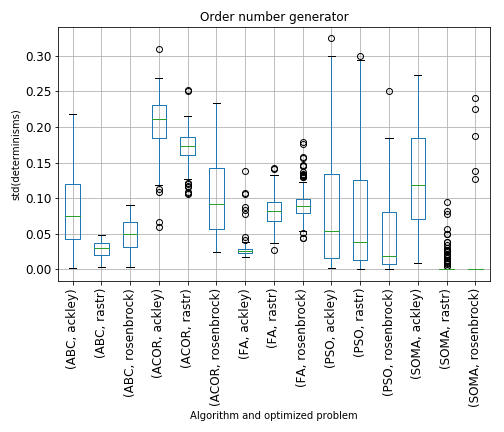}
            \includegraphics[scale=0.325]{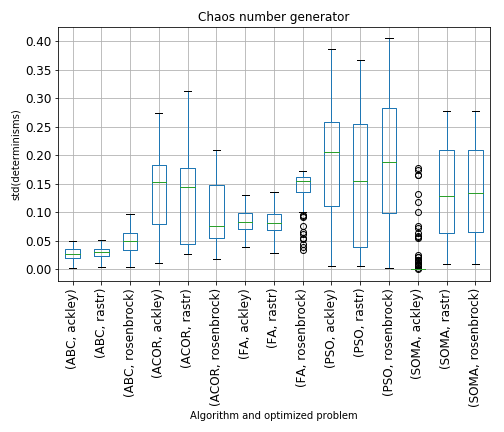}
        }
        \subfigure[]{
            \includegraphics[scale=0.325]{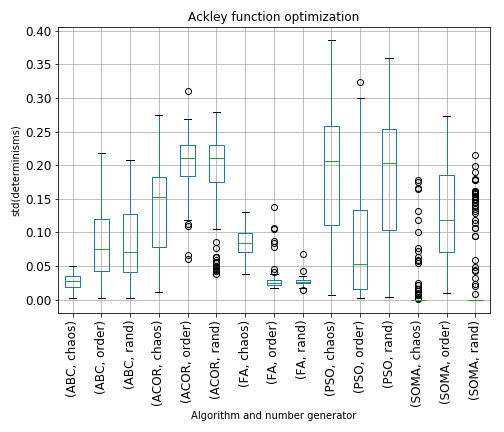}
            \includegraphics[scale=0.325]{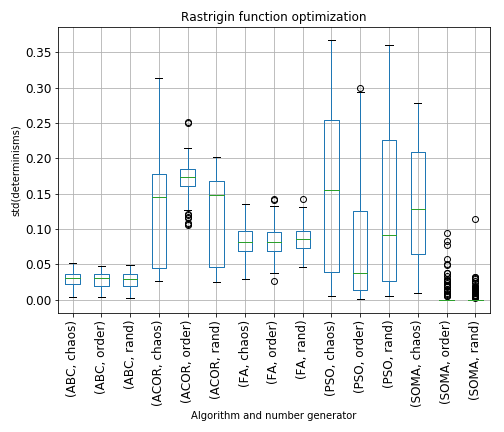}
            \includegraphics[scale=0.325]{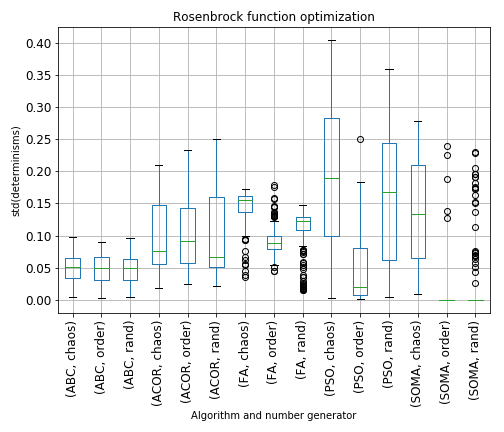}
        }
        \caption{Comparative visualization of mean distributions of standard deviations of determinisms of algorithm progresses. Sub-figures show the values separated on different criteria comparing their impact: (a) applied filtering based on number generator while subfilters were optimization algorithm and optmized function, (b) applied filtering based on optimized function while subfilters were optimization algorithm and number generator}
        \label{fig:viz_means}
    \end{figure*}
    
    \section{Results}
    
    The number of all algorithms executions was 200 and during them, the hyper-parameters were adjusted randomly in order to fairly examine the presence of phase transitions regardless of the optimization performance. The significance of phase transition $\tau$ was estimated by the standard deviation of the estimated complexity parameter (Determinism, Entropy and Kolmogorov complexity). The higher level of $\tau$ implies a higher level of fluctuating behavior of the optimization while minimal $\tau$ stood for a transition-less optimization. 
    
    In Fig.~\ref{fig:viz_means} various levels of phase transitions have been observed across all the available setups. From the given charts, there may be spotted those three defined factors affecting the significance of the phase transitions ($A,G,F$). Their impact is visually different, therefore our further examination is to reveal which of those affect the level of $\tau$ the most significantly. 
    
    Simply by filtering of one examined factor from the entire database and averaging its $\tau$ values on the given subset, we may estimate whether the optimization process is rather stable or not while applying this examined factor. To estimate how much one factor is affecting another, we need to perform another filtering on the given subset. This secondary filtering will reveal the conditional phase transition significance based on the given subfactor and it will show how this subfactor behaves on the given conditions. We started by filtering based on optimization algorithm $A$ and remaining factors for subselections were $G$ and $F$, therefore we were able to estimate how modification of $G$ and $F$ alters the behavior of $A$. Later we switched and as a main factor we selected $G$ and $F$ accordingly.
    
    \subsection{Selection by type of algorithm}
    
    Separation of the data set by algorithm factor $A$ results into five different subsets where we are able to examine the influence of number generator $G$ and optimization problem $F$ (see Figs.~\ref{tab_det}, \ref{tab_ent}, and \ref{tab_kc}). 
    Comparing the given setups, we may observe some differences among phase transition levels. It is difficult to estimate from these charts, whether the number generator is affecting the phase transitions more than the optimized function or vice versa. What is clearly visible, and will be estimated further as well, is a significant difference among algorithms which is caused by the way how they are designed to process the computation. Among algorithms with the rather higher mean of $\tau$, we may account PSO and $\text{ACO}_{\rm I\!R}$ especially due to values examined by Determinism and Kolmogorov complexity estimation. On the other hand, very stable behavior may be seen in cases of SOMA and ABC examinations.

\begin{table*}[!tbp]
            \caption{Determinism parameters: calculated averages of standard deviations of $\tau$ for the given setups to picture the amount of phase transition observed during optimization.}
        \subfigure[a][Artificial Bee Colony]{%
            \begin{tabular}{| l | c | c | c || c |}
Driver & ack. & ras. & ros. & all \\
\hline
chaos & 0.03 & 0.03 & 0.05 & 0.03 \\ 
order & 0.08 & 0.03 & 0.05 & 0.05 \\ 
rand & 0.08 & 0.03 & 0.05 & 0.05 \\ 
\hline
all  & 0.07 & 0.03 & 0.05 & \\
        \end{tabular}}
        \subfigure[b][Ant Colony]{
            \begin{tabular}{| l | c | c | c || c |}
Driver & ack. & ras. & ros. & all \\
\hline
chaos & 0.14 & 0.12 & 0.1 & 0.12 \\ 
order & 0.21 & 0.17 & 0.1 & 0.16 \\ 
rand & 0.19 & 0.12 & 0.1 & 0.14 \\ 
\hline
all  & 0.18 & 0.14 & 0.1 & \\
        \end{tabular}}
        \subfigure[][Particle Swarm]{
            \begin{tabular}{| l | c | c | c || c |}
Driver & ack. & ras. & ros. & all \\
\hline
chaos & 0.19 & 0.16 & 0.19 & 0.18 \\ 
order & 0.08 & 0.07 & 0.05 & 0.07 \\ 
rand & 0.18 & 0.13 & 0.16 & 0.16 \\ 
\hline
all  & 0.15 & 0.12 & 0.13 & \\
        \end{tabular}}
        \centering
        \subfigure[][Self organizing migrating]{
            \begin{tabular}{| l | c | c | c || c |}
Driver & ack. & ras. & ros. & all \\
\hline
chaos & 0.01 & 0.13 & 0.14 & 0.09 \\ 
order & 0.13 & 0.0 & 0.01 & 0.05 \\ 
rand & 0.02 & 0.0 & 0.02 & 0.01 \\ 
\hline
all  & 0.05 & 0.05 & 0.05 & \\
        \end{tabular}}
        \subfigure[][Firefly algorithm]{
            \begin{tabular}{| l | c | c | c || c |}
Driver & ack. & ras. & ros. & all \\
\hline
chaos & 0.08 & 0.08 & 0.14 & 0.1 \\ 
order & 0.03 & 0.08 & 0.09 & 0.08 \\ 
rand & 0.03 & 0.09 & 0.11 & 0.07 \\ 
\hline
all  & 0.05 & 0.08 & 0.11 & \\
        \end{tabular}}
        \label{tab_det}
    \end{table*}
    
    \begin{table*}[!tbp]
            \caption{Entropy parameter: calculated averages of standard deviations of $\tau$ for the given setups to picture the amount of phase transition observed during optimization.}
        \subfigure[a][Artificial Bee Colony]{%
            \begin{tabular}{| l | c | c | c || c |}
Driver & ack. & ras. & ros. & all \\
\hline
chaos & 0.14 & 0.16 & 0.12 & 0.14 \\ 
order & 0.21 & 0.16 & 0.12 & 0.16 \\ 
rand & 0.21 & 0.15 & 0.13 & 0.16 \\ 
\hline
all  & 0.19 & 0.16 & 0.12 & \\
        \end{tabular}}
        \subfigure[b][Ant Colony]{
            \begin{tabular}{| l | c | c | c || c |}
Driver & ack. & ras. & ros. & all \\
\hline
chaos & 0.55 & 0.56 & 0.34 & 0.49 \\ 
order & 0.86 & 0.71 & 0.33 & 0.64 \\ 
rand & 0.81 & 0.57 & 0.34 & 0.57 \\ 
\hline
all  & 0.74 & 0.61 & 0.34 & \\
        \end{tabular}}
        \subfigure[][Particle Swarm]{
            \begin{tabular}{| l | c | c | c || c |}
Driver & ack. & ras. & ros. & all \\
\hline
chaos & 0.48 & 0.44 & 0.48 & 0.47 \\ 
order & 0.4 & 0.18 & 0.16 & 0.26 \\ 
rand & 0.46 & 0.37 & 0.44 & 0.42 \\ 
\hline
all  & 0.44 & 0.33 & 0.35 & \\
        \end{tabular}}
        \centering
        \subfigure[][Self organizing migrating]{
            \begin{tabular}{| l | c | c | c || c |}
Driver & ack. & ras. & ros. & all \\
\hline
chaos & 0.05 & 0.56 & 0.57 & 0.38 \\ 
order & 0.51 & 0.03 & 0.02 & 0.19 \\ 
rand & 0.1 & 0.04 & 0.08 & 0.07 \\ 
\hline
all  & 0.21 & 0.2 & 0.22 & \\
        \end{tabular}}
        \subfigure[][Firefly algorithm]{
            \begin{tabular}{| l | c | c | c || c |}
Driver & ack. & ras. & ros. & all \\
\hline
chaos & 0.5 & 0.53 & 0.78 & 0.58 \\ 
order & 0.1 & 0.52 & 0.2 & 0.31 \\ 
rand & 0.08 & 0.51 & 0.35 & 0.31 \\ 
\hline
all  & 0.25 & 0.52 & 0.37 & \\
        \end{tabular}}
        \label{tab_ent}
    \end{table*}
    
    \begin{table*}[!tbp]
            \caption{Kolmogorov complexity parameter: calculated averages of standard deviations of $\tau$ for the given setups to picture the amount of phase transition observed during optimization.}
        \subfigure[a][Artificial Bee Colony]{%
            \begin{tabular}{| l | c | c | c || c |}
Driver & ack. & ras. & ros. & all \\
\hline
chaos & 0.14 & 0.17 & 0.23 & 0.18 \\ 
order & 0.24 & 0.16 & 0.23 & 0.21 \\ 
rand & 0.25 & 0.16 & 0.23 & 0.21 \\ 
\hline
all  & 0.21 & 0.16 & 0.23 & \\
        \end{tabular}}
        \subfigure[b][Ant Colony]{
            \begin{tabular}{| l | c | c | c || c |}
Driver & ack. & ras. & ros. & all \\
\hline
chaos & 0.8 & 0.59 & 0.75 & 0.72 \\ 
order & 1.36 & 0.98 & 0.71 & 1.01 \\ 
rand & 1.3 & 0.71 & 0.61 & 0.87 \\ 
\hline
all  & 1.15 & 0.76 & 0.69 & \\
        \end{tabular}}
        \subfigure[][Particle Swarm]{
            \begin{tabular}{| l | c | c | c || c |}
Driver & ack. & ras. & ros. & all \\
\hline
chaos & 1.03 & 1.1 & 1.0 & 1.04 \\ 
order & 0.55 & 0.56 & 0.57 & 0.56 \\ 
rand & 1.13 & 1.11 & 0.93 & 1.06 \\ 
\hline
all  & 0.91 & 0.92 & 0.83 & \\
        \end{tabular}}
        \centering
        \subfigure[][Self organizing migrating]{
            \begin{tabular}{| l | c | c | c || c |}
Driver & ack. & ras. & ros. & all \\
\hline
chaos & 0.1 & 0.88 & 0.88 & 0.62 \\ 
order & 0.89 & 0.02 & 0.03 & 0.31 \\ 
rand & 0.16 & 0.1 & 0.13 & 0.13 \\ 
\hline
all  & 0.38 & 0.33 & 0.35 & \\
        \end{tabular}}
        \subfigure[][Firefly algorithm]{
            \begin{tabular}{| l | c | c | c || c |}
Driver & ack. & ras. & ros. & all \\
\hline
chaos & 0.9 & 0.9 & 0.96 & 0.92 \\ 
order & 0.28 & 0.86 & 0.39 & 0.56 \\ 
rand & 0.27 & 0.9 & 0.95 & 0.71 \\ 
\hline
all  & 0.52 & 0.88 & 0.73 & \\
        \end{tabular}}
        \label{tab_kc}
    \end{table*}

    From the Tables \ref{tab_det}, \ref{tab_ent}, and \ref{tab_kc} it is not clear which of the secondary factors has higher influence. In cases of ABC, $\text{ACO}_\mathbb{R}$ and FA, the optimized function affects $\tau$ more significantly than the number generator. Differences of $\tau$ means are much higher based on $G$ compare to $F$. In other cases (PSO and SOMA) the much higher influence is obtained altering the number generator, while optimization seems not to be so sensitive on changing the optimization function. These findings were spotted in all three examined metrics $M$. The difference between those subfactors is rather small, but still, we may observe that changing the optimization function may change the phase transition significance more likely than changing the number generator.

    \subsection{Selection by type of number generator}

    In this second view on the result data, we will filter based on the number generator $G$ at first and than as a subfactors to compare, we will use the type of algorithm $A$ and optimized function $F$. Results are depicted in Tables \ref{tab_det_gen}, \ref{tab_ent_gen} and \ref{tab_kc_gen}. The differences on $\tau$ mean are much higher on algorithm based filtering compare to the optimized function based filtering. This simply implies that optimization procedure alters its phase transitions significance based on the kind of applied algorithm rather than optimized function. This observation was spotted in all kinds of examined metrics.

    \begin{table*}[!tbp]
            \caption{Determinism parameter: calculated averages of standard deviations of $\tau$ for the given setups to picture the amount of phase transition observed during optimization.}
        \subfigure[a][Random number generator]{%
            \begin{tabular}{| l | c | c | c || c |}
Driver & ack. & ras. & ros. & all \\
\hline
ABC & 0.08 & 0.03 & 0.05 & 0.05 \\ 
$\text{ACO}_\mathbb{R}$ & 0.19 & 0.12 & 0.1 & 0.14 \\ 
FA & 0.03 & 0.09 & 0.11 & 0.07 \\ 
PSO & 0.18 & 0.13 & 0.16 & 0.16 \\ 
SOMA & 0.02 & 0.01 & 0.02 & 0.01 \\ 
\hline
all  & 0.1 & 0.07 & 0.08 & \\
        \end{tabular}}
        \subfigure[b][Order number generator]{
            \begin{tabular}{| l | c | c | c || c |}
Driver & ack. & ras. & ros. & all \\
\hline
ABC & 0.08 & 0.03 & 0.05 & 0.05 \\ 
$\text{ACO}_\mathbb{R}$ & 0.21 & 0.17 & 0.1 & 0.16 \\ 
FA & 0.03 & 0.08 & 0.09 & 0.08 \\ 
PSO & 0.08 & 0.07 & 0.05 & 0.07 \\ 
SOMA & 0.13 & 0.0 & 0.01 & 0.05 \\ 
\hline
all  & 0.11 & 0.07 & 0.06 & \\
        \end{tabular}}
        \subfigure[][Chaos number generator]{
            \begin{tabular}{| l | c | c | c || c |}
Driver & ack. & ras. & ros. & all \\
\hline
ABC & 0.03 & 0.03 & 0.05 & 0.03 \\ 
$\text{ACO}_\mathbb{R}$ & 0.14 & 0.12 & 0.1 & 0.12 \\ 
FA & 0.08 & 0.08 & 0.14 & 0.1 \\ 
PSO & 0.19 & 0.16 & 0.19 & 0.18 \\ 
SOMA & 0.01 & 0.13 & 0.14 & 0.09 \\ 
\hline
all  & 0.09 & 0.11 & 0.12 & \\
        \end{tabular}}
        \label{tab_det_gen}
    \end{table*}

    \begin{table*}[!tbp]
            \caption{Entropy parameter: calculated averages of standard deviations of $\tau$ for the given setups to picture the amount of phase transition observed during optimization.}
        \subfigure[a][Random number generator]{%
            \begin{tabular}{| l | c | c | c || c |}
Driver & ack. & ras. & ros. & all \\
\hline
ABC & 0.21 & 0.15 & 0.13 & 0.16 \\ 
$\text{ACO}_\mathbb{R}$ & 0.81 & 0.57 & 0.34 & 0.57 \\ 
FA & 0.08 & 0.51 & 0.35 & 0.31 \\ 
PSO & 0.46 & 0.37 & 0.44 & 0.42 \\ 
SOMA & 0.1 & 0.04 & 0.08 & 0.07 \\ 
\hline
all  & 0.33 & 0.33 & 0.26 & \\
        \end{tabular}}
        \subfigure[b][Order number generator]{
            \begin{tabular}{| l | c | c | c || c |}
Driver & ack. & ras. & ros. & all \\
\hline
ABC & 0.21 & 0.16 & 0.12 & 0.16 \\ 
$\text{ACO}_\mathbb{R}$ & 0.86 & 0.71 & 0.33 & 0.64 \\ 
FA & 0.1 & 0.52 & 0.2 & 0.31 \\ 
PSO & 0.4 & 0.18 & 0.16 & 0.26 \\ 
SOMA & 0.51 & 0.03 & 0.02 & 0.19 \\ 
\hline
all  & 0.45 & 0.32 & 0.17 & \\
        \end{tabular}}
        \subfigure[][Chaos number generator]{
            \begin{tabular}{| l | c | c | c || c |}
Driver & ack. & ras. & ros. & all \\
\hline
ABC & 0.14 & 0.16 & 0.12 & 0.14 \\ 
$\text{ACO}_\mathbb{R}$ & 0.55 & 0.56 & 0.34 & 0.49 \\ 
FA & 0.5 & 0.53 & 0.78 & 0.58 \\ 
PSO & 0.48 & 0.44 & 0.48 & 0.47 \\ 
SOMA & 0.05 & 0.56 & 0.57 & 0.38 \\ 
\hline
all  & 0.34 & 0.44 & 0.42 & \\
        \end{tabular}}
        \label{tab_ent_gen}
    \end{table*}

    \begin{table*}[!tbp]
            \caption{Kolmogorov complexity parameter: calculated averages of standard deviations of $\tau$ for the given setups to picture the amount of phase transition observed during optimization.}
        \subfigure[a][Random number generator]{%
            \begin{tabular}{| l | c | c | c || c |}
Driver & ack. & ras. & ros. & all \\
\hline
ABC & 0.25 & 0.16 & 0.23 & 0.21 \\ 
$\text{ACO}_\mathbb{R}$ & 1.3 & 0.71 & 0.61 & 0.87 \\ 
FA & 0.27 & 0.9 & 0.95 & 0.71 \\ 
PSO & 1.13 & 1.11 & 0.93 & 1.06 \\ 
SOMA & 0.16 & 0.1 & 0.13 & 0.13 \\ 
\hline
all  & 0.62 & 0.59 & 0.57 & \\
        \end{tabular}}
        \subfigure[b][Order number generator]{
            \begin{tabular}{| l | c | c | c || c |}
Driver & ack. & ras. & ros. & all \\
\hline
ABC & 0.24 & 0.16 & 0.23 & 0.21 \\ 
$\text{ACO}_\mathbb{R}$ & 1.36 & 0.98 & 0.71 & 1.01 \\ 
FA & 0.28 & 0.86 & 0.39 & 0.56 \\ 
PSO & 0.55 & 0.56 & 0.57 & 0.56 \\ 
SOMA & 0.89 & 0.02 & 0.03 & 0.31 \\ 
\hline
all  & 0.71 & 0.52 & 0.39 & \\
        \end{tabular}}
        \subfigure[][Chaos number generator]{
            \begin{tabular}{| l | c | c | c || c |}
Driver & ack. & ras. & ros. & all \\
\hline
ABC & 0.14 & 0.17 & 0.23 & 0.18 \\ 
$\text{ACO}_\mathbb{R}$ & 0.8 & 0.59 & 0.75 & 0.72 \\ 
FA & 0.9 & 0.9 & 0.96 & 0.92 \\ 
PSO & 1.03 & 1.1 & 1.0 & 1.04 \\ 
SOMA & 0.1 & 0.88 & 0.88 & 0.62 \\ 
\hline
all  & 0.59 & 0.71 & 0.74 & \\
        \end{tabular}}
        \label{tab_kc_gen}
    \end{table*}
    
    SOMA with ABC appeared as the most stable having the lowest values of average differences of complexity parameters, while PSO and $\text{ACO}_\mathbb{R}$ performed the exact opposite indicating the much higher presence of phase transitions in this algorithm. FA was performing rather transitions occurring computations mostly on average of the observed algorithms. 
    
\subsection{Selection by type of optimized function}

    The filter based on the optimized function $F$ only confirms the previously observed findings, but this time the compared subfactors are the type of algorithm $A$ and number generator $G$. Results are depicted in Tables \ref{tab_det_op}, \ref{tab_ent_op}, and \ref{tab_kc_op}. The differences on $\tau$ mean are significantly higher on algorithm based filtering compare to the number generator based filtering. This again implies, that optimization procedure alters its phase transitions significance due to the kind of applied algorithm more likely than optimized function. This observation was spotted in all kinds of examined metrics.

    \begin{figure*}[!tbp]
        \centering
        \subfigure[]{\includegraphics[scale=0.375]{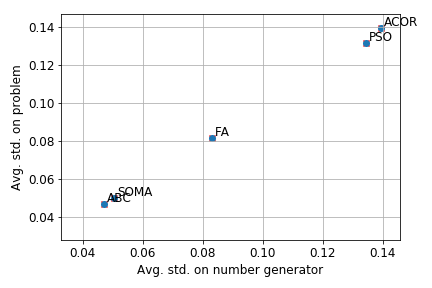}}
        \subfigure[]{\includegraphics[scale=0.375]{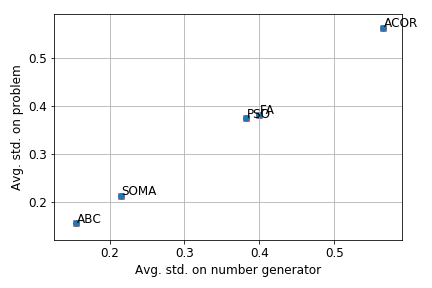}}
        \subfigure[]{\includegraphics[scale=0.375]{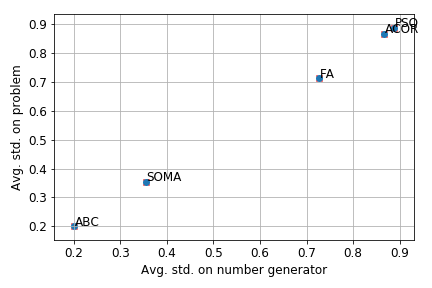}}
        \caption{Estimated sensitivity of the optimization algorithms' phase transitions towards used kind of fitness function and the number generator.(a) Standard deviation values of Determinism, (b) Standard deviation values of entropy and (c) Standard deviation values of Kolmogorov complexity.}
        \label{alg_map}
    \end{figure*}
    
    Due to our finding, the third examination of phase transitions was performed on the results filtered by algorithms' kind. We averaged the impact of the other factors for each algorithm to measure how they change alters the algorithms' behavior. The results are depicted in Fig.~\ref{alg_map} where we can clearly observe that PSO, $\text{ACO}_\mathbb{R}$ and FA are the groups of algorithms with the higher level of phase transitions while SOMA and ABC are representatives of rather phase-stable optimization approaches. These results were obtained similarly on all examined complexity measures with visible correlation among them.

        \begin{table*}[!tbp]
            \caption{Determinism parameter: calculated averages of standard deviations of $\tau$ for the given setups to picture the amount of phase transition observed during optimization.}
        \subfigure[a][Ackley function]{%
            \begin{tabular}{| l | c | c | c || c |}
Driver & chaos & order & rand & all \\
\hline
ABC & 0.03 & 0.08 & 0.08 & 0.07 \\ 
$\text{ACO}_\mathbb{R}$ & 0.14 & 0.21 & 0.19 & 0.18 \\ 
FA & 0.08 & 0.03 & 0.03 & 0.05 \\ 
PSO & 0.19 & 0.08 & 0.18 & 0.15 \\ 
SOMA & 0.01 & 0.13 & 0.02 & 0.05 \\ 
\hline
all  & 0.09 & 0.11 & 0.1 & \\
        \end{tabular}}
        \subfigure[b][Rastrigin function]{
            \begin{tabular}{| l | c | c | c || c |}
Driver & chaos & order & rand & all \\
\hline
ABC & 0.03 & 0.03 & 0.03 & 0.03 \\ 
$\text{ACO}_\mathbb{R}$ & 0.12 & 0.17 & 0.12 & 0.14 \\ 
FA & 0.08 & 0.08 & 0.09 & 0.08 \\ 
PSO & 0.16 & 0.07 & 0.13 & 0.12 \\ 
SOMA & 0.13 & 0.01 & 0.01 & 0.05 \\ 
\hline
all  & 0.11 & 0.07 & 0.07 & \\
        \end{tabular}}
        \subfigure[][Rosenbrock function]{
            \begin{tabular}{| l | c | c | c || c |}
Driver & chaos & order & rand & all \\
\hline
ABC & 0.05 & 0.05 & 0.05 & 0.05 \\ 
$\text{ACO}_\mathbb{R}$ & 0.1 & 0.1 & 0.1 & 0.1 \\ 
FA & 0.14 & 0.09 & 0.11 & 0.11 \\ 
PSO & 0.19 & 0.05 & 0.16 & 0.13 \\ 
SOMA & 0.14 & 0.01 & 0.02 & 0.05 \\ 
\hline
all  & 0.12 & 0.06 & 0.08 & \\
        \end{tabular}}
        \label{tab_det_op}
    \end{table*}

            \begin{table*}[!tbp]
            \caption{Enropy parameter: calculated averages of standard deviations of $\tau$ for the given setups to picture the amount of phase transition observed during optimization.}
        \subfigure[a][Ackley function]{%
            \begin{tabular}{| l | c | c | c || c |}
Driver & chaos & order & rand & all \\
\hline
ABC & 0.14 & 0.21 & 0.21 & 0.19 \\ 
$\text{ACO}_\mathbb{R}$ & 0.55 & 0.86 & 0.81 & 0.74 \\ 
FA & 0.5 & 0.1 & 0.08 & 0.25 \\ 
PSO & 0.48 & 0.4 & 0.46 & 0.44 \\ 
SOMA & 0.05 & 0.51 & 0.1 & 0.21 \\ 
\hline
all  & 0.34 & 0.45 & 0.33 & \\
        \end{tabular}}
        \subfigure[b][Rastrigin function]{
            \begin{tabular}{| l | c | c | c || c |}
Driver & chaos & order & rand & all \\
\hline
ABC & 0.16 & 0.16 & 0.15 & 0.16 \\ 
$\text{ACO}_\mathbb{R}$ & 0.56 & 0.71 & 0.57 & 0.61 \\ 
FA & 0.53 & 0.52 & 0.51 & 0.52 \\ 
PSO & 0.44 & 0.18 & 0.37 & 0.33 \\ 
SOMA & 0.56 & 0.03 & 0.04 & 0.2 \\ 
\hline
all  & 0.44 & 0.32 & 0.33 & \\
        \end{tabular}}
        \subfigure[][Rosenbrock function]{
            \begin{tabular}{| l | c | c | c || c |}
Driver & chaos & order & rand & all \\
\hline
ABC & 0.12 & 0.12 & 0.13 & 0.12 \\ 
$\text{ACO}_\mathbb{R}$ & 0.34 & 0.33 & 0.34 & 0.34 \\ 
FA & 0.78 & 0.2 & 0.35 & 0.37 \\ 
PSO & 0.48 & 0.16 & 0.44 & 0.35 \\ 
SOMA & 0.57 & 0.02 & 0.08 & 0.22 \\ 
\hline
all  & 0.42 & 0.17 & 0.26 & \\
        \end{tabular}}
        \label{tab_ent_op}
    \end{table*}

            \begin{table*}[!tbp]
            \caption{Kolmogorov complexity parameter: calculated averages of standard deviations of $\tau$ for the given setups to picture the amount of phase transition observed during optimization.}
        \subfigure[a][Ackley function]{%
            \begin{tabular}{| l | c | c | c || c |}
Driver & chaos & order & rand & all \\
\hline
ABC & 0.14 & 0.24 & 0.25 & 0.21 \\ 
$\text{ACO}_\mathbb{R}$ & 0.8 & 1.36 & 1.3 & 1.15 \\ 
FA & 0.9 & 0.28 & 0.27 & 0.52 \\ 
PSO & 1.03 & 0.55 & 1.13 & 0.91 \\ 
SOMA & 0.1 & 0.89 & 0.16 & 0.38 \\ 
\hline
all  & 0.59 & 0.71 & 0.62 & \\
        \end{tabular}}
        \subfigure[b][Rastrigin function]{
            \begin{tabular}{| l | c | c | c || c |}
Driver & chaos & order & rand & all \\
\hline
ABC & 0.17 & 0.16 & 0.16 & 0.16 \\ 
$\text{ACO}_\mathbb{R}$ & 0.59 & 0.98 & 0.71 & 0.76 \\ 
FA & 0.9 & 0.86 & 0.9 & 0.88 \\ 
PSO & 1.1 & 0.56 & 1.11 & 0.92 \\ 
SOMA & 0.88 & 0.02 & 0.1 & 0.33 \\ 
\hline
all  & 0.71 & 0.52 & 0.59 & \\
        \end{tabular}}
        \subfigure[][Rosenbrock function]{
            \begin{tabular}{| l | c | c | c || c |}
Driver & chaos & order & rand & all \\
\hline
ABC & 0.23 & 0.23 & 0.23 & 0.23 \\ 
$\text{ACO}_\mathbb{R}$ & 0.75 & 0.71 & 0.61 & 0.69 \\ 
FA & 0.96 & 0.39 & 0.95 & 0.73 \\ 
PSO & 1.0 & 0.57 & 0.93 & 0.83 \\ 
SOMA & 0.88 & 0.03 & 0.13 & 0.35 \\ 
\hline
all  & 0.74 & 0.39 & 0.57 & \\
        \end{tabular}}
        \label{tab_kc_op}
    \end{table*}
    
    \section{Conclusions}

The varying instability of swarm optimization behavior was examined in these experiments in a slightly larger scale comparing to our first initial study \cite{vantuch2018phase}. 

Five swarm-intelligence based optimization algorithms were examined in nine different setups based on three different number generators and three different optimized functions. The main motivation was to compare which factor most likely affects the amount of phase transitions. From our simulations, it is clearly visible that the type of optimization algorithm is the key factor affecting the significance of phase transitions. The remaining factors were also altering this phenomenon significantly but in a much smaller scale. 

The last comparison only underlines our conclusions. Algorithms were depicted in Fig.~\ref{alg_map} where the sensitivity on the number generator (the average standard deviation on all number generators) was in all cases very close to the sensitivity on the fitness function (the average standard deviation on all fitness functions), while differences among the algorithms were very significant. All three complexity measures confirmed this observation with a slight visible correlation.

Our future work has to examine whether the phase transitions are beneficial for the convergence and which algorithm is using them this way, because otherwise, they may perform only disruptive element which is necessary to minimize. On the other hand, our results sometimes returned an outlier observations (behavior of some algorithm changed too much or not at all) which may be caused by another, not considered factor. Our future study will consider the examination of initial population distribution on the phase transition significance as well.

    \section*{ACKNOWLEDGMENT}
    \sloppy 
    This paper was supported by the following projects: LO1404: Sustainable development of ENET Centre; SP2019/28 and SGS 2019/137 Students Grant Competition and the Project LTI17023 ``Energy Research and Development Information Centre of the Czech Republic'' funded by Ministry of Education, Youth and Sports of the Czech Republic, program INTER-EXCELLENCE, subprogram INTER-INFORM, and DFG projects MA4759/8 and MA4759/9.

    \bibliographystyle{IEEEtran}
    \bibliography{refs}

\end{document}